\documentstyle[12pt]{article}
\begin{document}
\begin{center}
{\Large\bf Relativistic Stationary Schr$\ddot{\rm o}$dinger Equation for Many
Particle System and Its Applications}\\
\vspace{1cm}
{Guang-Jiong Ni$^{*}$\footnotetext{E-mail:
$^{*}$gjni@fudan.ihep.ac.cn}}
\end{center}
\centerline{\it Department of Physics,
Fudan University, Shanghai 200433, China}
\vspace{1cm}
\begin{center}
\begin{minipage}{128mm}
\baselineskip=12pt
\centerline{ABSTRACT}
\vspace{0.5cm}

Basing on the fundamental symmetry that the space-time inversion is
equavalent to particle-antiparticle transformation, a relativistic
modification on the stationary Schr$\ddot{\rm o}$dinger
equation for many-particle
system is made. The eigenvalue in the center of mass system is no longer
equal to the negative of binding energy simply. The possible
applications in various fields (e.g. the model of quarkonium) are
discussed.

\vspace{0.5cm}
\end{minipage}
\end{center}

\newpage

As is well known in quantum mechanics, for a system composed of two
particles with mass $m_1$ and $m_2$ and interaction potential
$V(|\vec{r_1}-\vec{r_2}|)$, ($\vec{r_1}$ and $\vec{r_2}$ being the
coordinates in laboratory (L) system), after introducing the coordinate
of center of mass (CM)
$\vec{R}=\frac{1}{M}(m_{1}\vec{r_1}+m_{2}\vec{r_{2}})$,
($M=m_{1}+m_{2}$), and relative coordinate
$\vec{r}=\vec{r_2}-\vec{r_1}$, one easily obtain the stationary
Schr$\ddot{\rm o}$dinger equation
\begin{equation}
[-\frac{\hbar^{2}}{2\mu}\nabla_{\vec{r}}^{2}+V(\vec{r})]\psi(\vec{r})
=\epsilon\psi(\vec{r})
\end{equation}
with reduced mass $\mu=m_{1}m_{2}/M$ and eigenvalue
\begin{equation}
\epsilon=E-\frac{P^2}{2M}-Mc^2
\end{equation}
where $E$ is the total energy of system in L system and $P=|\vec{P}|$ is
the momentum of CM. When $P=0$, $\epsilon=-B<0$, $B$ is called as the
binding energy of system.

Usually, it is said that the above ``nonrelativistic approximation'' is
good for the case of $P^{2}/(2M)$ being small. However, this is
doubtful. One can always set the laboratory into motion to render $P$
arbitrary large. Then the accuracy of calculation in CM system would
rely on the motion of external world ! Does it respect to the
``principle of relativity'' in special relativity (SR) ? The reason why
one can not isolate his calculation from the external world is the
following. The summation of $\epsilon$ with $P^{2}/(2M)$ is made along a
straight line in Eq.(2) whereas the correct relation in SR is that of a
right triangle:
\begin{equation}
E^{2}=P^{2}c^{2}+(Mc^{2}-B)^2
\end{equation}

Hence, the statement that ``the total energy E is equal to the sum of
kinetic energy ($P^{2}/2M$) of CM and the internal energy ($Mc^{2}-B$)''
is not rigorous. One should say that ``$E$ is equal to the square root
of the sum of square of kinetic energy ($Pc$) of CM and that of internal
energy ''. Now the problem is ``how to modify the Eq.(1) for meeting the
requirement of Eq.(3) ?''

It is said in ancient China that ``to gain new insights through
restudying old material''. Let us restudy an alternative derivation of
Klein-Gordon (KG) equation. Consider a spinless particle with rest mass
$m_0$ being in free motion. Then its wave function $\theta(\vec{x},t)$
is described by a ``nonrelativistic quantum equation'':
\begin{equation}
i\hbar\frac{\partial}{\partial t}\theta(\vec{x},t)=
m_{0}c_{1}^{2}\theta(\vec{x},t)
-\frac{\hbar^{2}}{2m_0}\nabla^{2}\theta(\vec{x},t)
\end{equation}
Here, we add a term of rest energy $m_{0}c_{1}^2$ with $c_{1}$ being
merely a (unfixed yet) constant with dimension of velocity. The next
crucial step is assuming that inside a particle state $\theta$, there is
always a hiding antiparticle state $\chi(\vec{x},t)$. $\theta$ and
$\chi$ are coupled together via motion. Instead of Eq.(4), we should
have a simultaneous equation as follows:
\begin{equation}
\left \{
\begin{array}{l}
i\hbar\frac{\partial}{\partial
t}\theta=m_{0}c_{1}^{2}\theta-\frac{\hbar^{2}}{2m_0}\nabla^{2}\theta
-\frac{\hbar^{2}}{2m_0}\nabla^{2}\chi \\
i\hbar\frac{\partial}{\partial
t}\chi=-m_{0}c_{1}^{2}\chi+\frac{\hbar^{2}}{2m_0}\nabla^{2}\chi
+\frac{\hbar^{2}}{2m_0}\nabla^{2}\theta
\end{array}
\right.
\end{equation}
The guiding rule for establishing the Eq.(5) is a basic symmetry. It is
invariant under the space-time inversion $\vec{x}\rightarrow -\vec{x}$,
$t\rightarrow -t$ and transformation:
\begin{equation}
\chi(\vec{x},t)=\theta(-\vec{x},-t)
\end{equation}
For solving Eq.(5) in general, we use the ansatz
\begin{equation}
\left \{
\begin{array}{l}
\theta=(\phi+i\frac{\hbar}{m_{0}c_{1}^2}\dot{\phi}) \\
\chi=(\phi-i\frac{\hbar}{m_{0}c_{1}^2}\dot{\phi})
\end{array}
\right.
\end{equation}
and obtain the K-G equation
\begin{equation}
(\frac{1}{c_{1}^{2}}\frac{\partial^2}{\partial t^{2}}-\nabla^{2}
+\frac{m_{0}^{2}c_{1}^2}{\hbar^2})\phi(\vec{x},t)=0
\end{equation}
Its plane wave solution
\begin{equation}
\phi(\vec{x},t)=exp[i(\vec{p}\cdot\vec{x}-Et)/\hbar]
\end{equation}
leads directly to
\begin{equation}
E^{2}=\vec{p}^{2}c_{1}^{2}+m_{0}^{2}c_{1}^{4}
\end{equation}
For clarifying the meaning of $c_{1}$, we look at the velocity ($v$) of
particle which is equal to the group velocity ($v_{g}$) of de'Broglie
wave:
\begin{equation}
v=v_{g}=\frac{d\omega}{dk}=\frac{dE}{dp}=pc_{1}^{2}/E,~~ (p=|\vec{p}|)
\end{equation}
where the quantum relations $E=\hbar\omega$ and $p=hk$ have been used.

The inertial mass is defined as
\begin{equation}
m=\frac{p}{v}=p/(\frac{dE}{dp})=\frac{1}{2}\frac{d}{dE}\vec{p}^{2}
\end{equation}
Combining Eqs.(10)-(12), we arrive at
\begin{equation}
E=mc_{1}^2
\end{equation}
\begin{equation}
m=m_{0}/\sqrt{1-\frac{v^2}{c_{1}^{2}}}
\end{equation}
as expected. But here $c$ has the meaning as the limiting speed of
particle, its value is obtained from the measurement on the $\pi$ meson
beam and is coinciding with the speed of light, $c$:
\begin{equation}
c_{1}=c=3\times 10^{10} {\rm cm/sec}
\end{equation}
In the above derivation of K-G equation (a reversed version of that in
Ref.[1]), we start from $E|_{v=0}=m_{0}c_{1}^{2}$ and get eventually the
mass energy relation $E=mc_{1}^{2}$. The proof bears some resemblance to
the ``inductive method'' in mathematics. However, the important thing is
injecting into the proof a ``relativistic principle'', i.e.,
the basic symmetry (6), which plays the role of ``hormone''
for activation of mass from $m_0$ into $m$.

The symmetry (6) is discussed generally as a statement that ``the
space-time inversion is equavalent to particle-antiparticle
transformation'' in Refs. [2-3]. In our opinion, it is a natrual
postulate after we learn carefully from the development of physics
since the discovery of parity violation [4,5] and the observation of
Schwinger et al. [6,7].

We are now in a position to generalize the above derivation to two
particle case as in Eq.(1). Denoting
$\theta=\theta(\vec{r_1},\vec{r_2},t)$ and
$\chi=\chi(\vec{r_1},\vec{r_2},t)$ the particle and corresponding
``antiparticle'' state again, then instead of Eq.(5), we can write down
the following simultaneous equation:
\begin{equation}
\left \{
\begin{array}{l}
i\hbar\frac{\partial\theta}{\partial t}=Mc^{2}\theta
-(\frac{\hbar^{2}}{2m_1}\nabla_{\vec{r_1}}^{2} +
\frac{\hbar^{2}}{2m_2}\nabla_{\vec{r_2}}^{2})(\theta+\chi)
+V(|\vec{r_1}-\vec{r_2}|)(\theta+\chi) \\
i\hbar\frac{\partial\chi}{\partial t}=-Mc^{2}\chi
+(\frac{\hbar^{2}}{2m_1}\nabla_{\vec{r_1}}^{2} +
\frac{\hbar^{2}}{2m_2}\nabla_{\vec{r_2}}^{2})(\theta+\chi)
-V(|\vec{r_1}-\vec{r_2}|)(\theta+\chi)
\end{array}
\right.
\end{equation}
which still respect to the symmetry:
\begin{equation}
\chi(\vec{r_1},\vec{r_2},t)=\theta(-\vec{r_1},-\vec{r_2},-t)
\end{equation}
As Eq.(7), we set
\begin{equation}
\theta=\Phi+i\frac{\hbar}{Mc^2}\dot{\Phi},~~
\chi=\Phi-i\frac{\hbar}{Mc^2}\dot{\Phi}
\end{equation}
with $\Phi(\vec{r_1},\vec{r_2},t)\rightarrow\Phi(\vec{R},\vec{r},t)$
obeying the equation:
\begin{equation}
\ddot{\Phi}-c^{2}\nabla_{\vec{R}}^{2}\Phi
-c^{2}\frac{M}{\mu}\nabla_{\vec{r}}^{2}\Phi
+\frac{1}{\hbar^{2}}(M^{2}c^{4}+2VMc^{2})\Phi=0
\end{equation}
Factorizing the solution as
\begin{equation}
\Phi(\vec{R},\vec{r},t)=e^{i\vec{P}\cdot\vec{R}/\hbar}
e^{-iEt/\hbar}\psi(\vec{r})
\end{equation}
and substituting it into Eq.(19), we arrive at
\begin{equation}
\left \{
\begin{array}{l}
[-\frac{\hbar^2}{2\mu}\nabla_{\vec{r}}^{2}+V(r)]\psi(\vec{r})
=\epsilon\psi(\vec{r}) \\
\epsilon=\frac{1}{2Mc^{2}}(E^{2}-M^{2}c^{4}-P^{2}c^{2})
\end{array}
\right.
\end{equation}

Note that the eigenvalue $\epsilon\neq E-Mc^2$ even when $P=0$. Comparing
Eq.(8) with Eq.(3), we find that the accurate relation between the
binding energy $B$ and $\epsilon$ reads
\begin{equation}
B=Mc^{2}[1-(1+\frac{2\epsilon}{Mc^{2}})^{1/2}]
\end{equation}
Only when $\frac{\epsilon}{Mc^{2}}<<1$, can one recover the
nonrelativistic approximation:
\begin{equation}
B\simeq -\epsilon
\end{equation}
In general case, one should use Eq.(22) or its equavalent (for $P=0$):
\begin{equation}
E|_{P=0}=[2Mc^{2}\epsilon+M^{2}c^{4}]^{1/2}
\end{equation}

It is easy to generalize the above consideration to many particle
($n\geq 3$) case. Denote the coordinates of $i$th particle with rest
mass $m_{i}$ are $\vec{r'}_{i}$ and $\vec{r}_{i}$ in L system and CM
system respectively. The coordinate of CM reads
$\vec{R}=\sum_{i=1}^{n}\frac{1}{M}m_{i}\vec{r'}_{i}$,
($M=\sum_{i=1}^{n}m_{i}$), while $\vec{r}_{i}=\vec{r'}_{i}-\vec{R}$
obeys the constraint:
\begin{equation}
\sum_{i=1}^{n}m_{i}\vec{r}_{i}=0
\end{equation}
Direct calculation leads to
\begin{equation}
\sum_{i=1}^{n}\frac{1}{m_{i}}\nabla_{\vec{r'}_{i}}^{2}=
\frac{1}{M}\nabla_{\vec{R}}^{2}
+\sum_{i=1}^{n}\frac{1}{m_{i}}\nabla_{\vec{r}_{i}}^{2}-
\frac{1}{M}(\sum_{i=1}^{n}\nabla_{\vec{r}_{i}})\cdot
(\sum_{j=1}^{n}\nabla_{\vec{r}_{j}})
\end{equation}
The third term in RHS can be discarded because in CM system
the total momentum
equals to zero in stationary state. Denoting
$\theta=\theta(\vec{r'}_{1},\vec{r'}_{2},\cdots,\vec{r'}_{n})$,
$\chi=\chi(\vec{r'}_{1},\vec{r'}_{2},\cdots,\vec{r'}_{n})$ and
introducing again the wave function
\begin{equation}
\Phi(\vec{R},\vec{r}_{i},t)=
=e^{i\vec{P}\cdot\vec{R}/\hbar}
e^{-iEt/\hbar}\psi(\vec{r}_{1},\vec{r}_{2},\cdots,\vec{r}_{n})
\end{equation}
we find
\begin{equation}
\left \{
\begin{array}{l}
[-\frac{\hbar^2}{2}\sum_{i=1}^{n}\frac{1}{m_i}
\nabla_{\vec{r}_{i}}^{2}+\sum_{i<j}^{n}V_{ij}(r_{ij})]\psi
=\epsilon\psi \\
\epsilon=\frac{1}{2Mc^{2}}(E^{2}-M^{2}c^{4}-P^{2}c^{2})
\end{array}
\right.
\end{equation}
Similar to Eq.(21). However, only ($n-1$) coordinates in $\vec{r}_{i}$
are independent due to the constraint (25).

Some remarks are in order:\\
1. There is something unreasonable in previous stationary Schr$\ddot{\rm
o}$dinger equation (1). The eigenvalue $\epsilon$ may fall
downward without lower bound.
This situation does occur for a singular potential
like $V(r)\sim \frac{-1}{r^2}$ (see Refs.[8-9]).
Now Eq.(21) has no this kind of worry. There is a minimum value for
$\epsilon$: $\epsilon_{\rm min}=-\frac{1}{2}Mc^{2}$, or $E_{\rm min}=0$.
Therefore, the solution of Eq.(21) may be viewed as a variational problem
and a lower bound exists for any variational procedure.\\
2. Let us compare Eq.(21) with Dirac equation. For an electron with
mass $m_{e}=m$ moving in the Coulomb field $V(r)=-Ze^{2}/(4\pi r)$ of
neucleus with mass $m_{N}\rightarrow\infty$, the total energy of
electron reads [10]:
\begin{equation}
E_{\rm D}=mc^{2}\{1+\frac{Z^{2}\alpha^{2}}
{[\sqrt{(j+1/2)^{2}-Z^{2}\alpha^{2}}+n^{'}]^{2}}\}^{-1/2}
\end{equation}
with $\alpha=\frac{e^{2}}{4\pi \hbar c}\simeq \frac{1}{137}$,
$j=\frac{1}{2},\frac{3}{2},\cdots$, $n^{'}=0,1,2,\cdots$. In terms of
the principal quantum number $n=n^{'}+(j+\frac{1}{2})$, one has
\begin{equation}
E_{\rm D}=mc^{2}[1-\frac{1}{2}\frac{(Z\alpha)^{2}}{n^2}
-\frac{1}{2}\frac{(Z\alpha)^{4}}{n^3}(\frac{1}{j+1/2}-\frac{3}{4n})
-\cdots ]
\end{equation}
For comparision, the rest energy of neucleus $m_{\rm N}c^{2}$ must be
substracted from the $E$ derived from Eq.(24). Denote
\begin{equation}
E_{\rm S}\equiv E-m_{\rm N}c^{2}=c^{2}(m+m_{N})
[1+\frac{2\epsilon}{(m+m_{\rm N})c^{2}}]^{1/2}-m_{\rm N}c^{2}
\end{equation}
where $\epsilon$ is well known as
$$
\epsilon=-\mu c^{2}\frac{Z^{2}\alpha^{2}}{2n^2},~~ (n=1,2,\cdots)
$$
\begin{equation}
E_{\rm S}=mc^{2}[1-\frac{1}{2}\frac{m_{N}}{(m+m_{N})}(\frac{Z\alpha}{n})^{2}
-\frac{1}{8}\frac{mm_{N}^2}{(m+m_{N})^3}(\frac{Z\alpha}{n})^{4}-\cdots]
\end{equation}
The difference between (30) and (32) is stemming from two reasons. In
Dirac equation the spin of electron is taken into account whereas in
Eq.(21) the finiteness of neucleas mass is important. However, both
equations are relativistic because the basic symmetry (6) is respected
(see Refs.[2-3]).\\
3. Being an improvement to Eq.(1), Eq.(21) or (28) brings some modification on
many problems in stationary states. Roughly speaking, for binding state
problem, the modification is very small in atomic physics ($<10^{-5}$),
it accounts for $10^{-3}$ in nuclear physics and may reach  $10^{-2}$ in
particle physics. For the high energy scattering problem, this
modification might be also important.\\
4. For concreteness, let us have a quick look at the potential model of
heavy quarkonium, $Q\bar{Q}$. Assume that the potential between $Q$ and
$\bar{Q}$ is of the linear type, $V(r)=ar$, with constant $a$ being
independent of quark mass $m$. For $S$ states,
the stationary equation is solved analytically with
eigenvalue
\begin{equation}
\epsilon_{n}=\lambda_{n}(\frac{a^{2}}{2\mu})^{1/3},~~ (n=1,2,\cdots)
\end{equation}
$\lambda_{n}$ being the zeros of Airy function [11]. So the energy
(mass) of $Q\bar{Q}$ reads from Eq.(24):
\begin{equation}
E_{n}=4\mu [1+\frac{\lambda_n}{2}(\frac{a^2}{2\mu^4})^{1/3}]^{1/2}
\end{equation}
with $\mu=\frac{m}{2}=\frac{M}{4}$. On the other hand, the previous
equation (1) with (2) yields
\begin{equation}
E_{n}^{'}=4\mu^{'}+\lambda_{n} ( \frac{a'^2}{2\mu^{'}} ) ^{1/3}
\end{equation}
Table 1 is a comparision between the measured energy $E^{\rm exp}_{n}$
of $S$ states in Upsilon $b\bar{b}$ [12] and $E_n$ or $E_{n}^{'}$. In either
case, the parameters $a$ and $\mu$ are adjusted so that $E_n$ or
$E_{n}^{'}$
is coinciding with $E^{\rm exp}_n$ for $n=1$ and $2$. the mass of
constituent quark $b$ is fitted as
\begin{equation}
m_{b}=2\mu=4.326 {\rm GeV}
\end{equation}
$$
a=0.4530 {\rm GeV ^2}
$$
from $E_n$ versus
\begin{equation}
m_{b}^{'}=2\mu^{'}=4.354 {\rm GeV}
\end{equation}
$$
a^{'}=0.3804 {\rm GeV ^2}
$$
from $E_{n}^{'}$. The general trend of $E_n$ for higher $n$ seems better
than that of $E_{n}^{'}$ as shown in the table.\\
5. Similar fitting procedure used for Charmonium $J/\psi=c\bar{c}$ by
Eq.(34) leads to
\begin{equation}
m_{c}=1.031 {\rm GeV}
\end{equation}
$$
a=0.4183 {\rm GeV ^2}
$$
whereas Eq.(35) yields
\begin{equation}
m_{c}^{'}=1.155 {\rm GeV}
\end{equation}
$$
a^{'}=0.2099 {\rm GeV ^2}
$$
6. If neglecting the dependence of constant $a$ in $V(r)=ar$ on the
quark mass, we may discuss the dependence of quarkonium mass on the
quark mass for the level with same quantum numbers. The Feynman-Hellmann
theorem for stationary Schr$\ddot{\rm o}dinger$ equation reads [13]
\begin{equation}
\frac{\partial\epsilon}{\partial\mu}=-\frac{1}{\mu}(\epsilon-\langle V
\rangle ) <0
\end{equation}
Now it is replaced by
\begin{equation}
\frac{\partial E}{\partial\mu}=-\frac{4}{E}(\epsilon-\langle V
\rangle ) +\frac{8\mu}{E}+\frac{E}{2\mu}
\end{equation}
The latter two terms in RHS are positive. Combining Eq.(41) further with
the virial theorem:
\begin{equation}
\epsilon-\langle V\rangle\equiv\langle T\rangle=\langle
\frac{1}{2}r\frac{dV}{dr}\rangle
\end{equation}
we get for $V(r)=ar$:
\begin{equation}
\frac{\partial
E}{\partial\mu}=\frac{1}{3}\frac{E}{\mu}+\frac{32}{3}\frac{\mu}{E}
\end{equation}
versus
\begin{equation}
\frac{\partial
E^{'}}{\partial\mu^{'}}=\frac{16}{3}-\frac{1}{3}\frac{E^{'}}{\mu^{'}}
\end{equation}
from $E^{'}=\epsilon+4\mu^{'}$.
Eq.(44) can easily be integrated as
\begin{equation}
E^{'}(\mu^{'})=4\mu^{'}+C^{'}{\mu^{'}}^{-1/3}
\end{equation}
whereas Eq.(43) can also be linearized by $E=\sqrt{y}$ and integrated as
\begin{equation}
E(\mu)=[16\mu^{2}+C\mu^{2/3}]^{1/2}
\end{equation}
As an interesting test, we use the experimental data of ground state
energy for Charmonim $E_{1}=3.097$ GeV to fix the value of $C$ in
Eq.(46) or $C^{'}$ in Eq.(45) (using the value of $\mu$ or $\mu^{'}$ in the
Eqs.(38) or (39) at the same time). Then the ground state of $b\bar{b}$
can be estimated as
\begin{equation}
E_{1}(b\bar{b})=9.420 {\rm GeV}
\end{equation}
or
\begin{equation}
E_{1}^{'}(b\bar{b})=9.214 {\rm GeV}
\end{equation}
versus
\begin{equation}
E_{1}^{\rm exp}(b\bar{b})=9.460 {\rm GeV}
\end{equation}
Similarily for the $2S$ state (3.686 GeV) of Charmonium
with other value of $C$ or $C^{'}$, we get
\begin{equation}
E_{2}(b\bar{b})=9.9562 {\rm GeV}
\end{equation}
\begin{equation}
E_{2}^{'}(b\bar{b})=9.5922 {\rm GeV}
\end{equation}
versus
\begin{equation}
E_{2}^{\rm exp}(b\bar{b})=10.023 {\rm GeV}
\end{equation}

The author would like to thank Su-qing Chen and Ji-feng Yang for
discussions. This work was supported in part by the NSF in China.

\newpage

\vspace*{1.5cm}
\noindent {\bf References}
\vspace*{4mm}

\noindent [1] H. Feshbach and F. M. H. Villars,
Rev. Mod. Phys. {\bf 30},24 (1958).

\noindent [2] G-j Ni and S-q Chen, preprint, Internet Hep-th/9508069.

\noindent [3] G-j Ni and S-q Chen, Journal of Fudan University (Natural
Science), {\bf 35}(3), 326, (1996).

\noindent [4] T. D. Lee and C. N. Yang, Phys. Rev. {\bf 104}, 254
(1956).

\noindent [5] C. S. Wu, E. Ambler, R. W. Hayward, D. D. Hoppes and R. P.
Hudson, Phys. Rev. {\bf 105}, 1413 (1957);
R. L. Garwin, L. M. Lederman and M. Weinrich,
Phys. Rev. {\bf 105}, 1415 (1957);
V. L. Telegdi and A. M.
Friedman, Phys. Rev. {\bf 105}, 1681 (1957).

\noindent [6] J. Schwinger, Proc. Nat. Acad. Sc. U.S. {\bf 44}, 223
(1958).

\noindent [7] E. J. Konopinski and H. M. Mahmaud, Phys. Rev. {\bf 92},
1045 (1953).

\noindent [8] L. D. Landau and E. M. Lifshitz, {\it Quantum Mechanics:
Non-relativistic Theory} (Pergamon Press, 1977), 114

\noindent [9] G-j Ni and S-q Chen, {\it Levinson Theorem, Anomaly and the
Phase Transition of Vacuum} (Shanghai Scientific \& Technical
Publishers, 1995), Chapter 2.

\noindent [10] J. D. Bjorken and S. D. Drell, {\it Relativistic Quantum
Mechanics} (McGraw-Hill Book Company, 1964), 55.

\noindent [11] J. J. Sakurai, {\it Modern Quantum Mechanics}
(The Benjamin / Cummings Publishing Company, 1985), 108;
S. Fl$\ddot{\rm u}$gge, {\it Practical Quantum Mechanics}
(Springer-Verlag, 1970), 101.

\noindent [12] Review of Particle Physics, Phys. Rev. D {\bf 54}, No.1 (1996).

\noindent [13] C. Quigg and J. L. Rosner, Phys. Rep. {\bf 56}, 167 (1979).

\begin{table}
\caption{The $S$ states of Upsilon $b\bar{b}$.}
\begin{tabular}{||c|c|c|c|c|c|c||}
\hline
\hline
n & 1 & 2 & 3 & 4 & 5 & 6\\
\hline
$E_{n}^{\rm exp}$ (GeV) & 9.46037 & 10.023 & 10.355 & 10.580 & 10.865 &
11.019\\
\hline
$E_{n}$ (GeV) & 9.46037 & 10.023 & 10.461 & 10.834 & 11.163 & 11.462\\
\hline
$E_{n}^{'}$ (GeV) & 9.46037 & 10.023 & 10.483 & 10.890 & 11.262 & 11.609\\
\hline
$\lambda_{n}$ & 2.338 & 4.088 & 5.521 & 6.787 & 7.944 & 9.023\\
\hline
\hline
\end{tabular}
\end{table}

\end{document}